\documentstyle[twoside,fleqn,espcrc2]{article}

\newcommand{\beq}{\begin{eqnarray}}
\newcommand{\eeq}{\end{eqnarray}}

\newcommand{\kbra}[1] { \left< #1 \right>}

\newcommand{\AmS}{{\protect\the\textfont2
  A\kern-.1667em\lower.5ex\hbox{M}\kern-.125emS}}

\title{
\vspace{-2cm}
\begin{flushright}
KANAZAWA 99-19\ \ \ \ \ \ \\
August 1999\ \ \ \ \ \ \  
\end{flushright}
\vspace*{-.5cm}
String tension and glueball masses of SU(2) QCD
from perfect action for monopoles and strings 
\thanks{presented by S. Kato}}
\author{
S. Fujimoto, 
\hspace{2mm}
\thanks{ E-mail address:kato@hep.s.kanazawa-u.ac.jp }
S. Kato, 
\hspace{2mm}
M. Murata
\hspace{2mm}
and \hspace{2mm}
T. Suzuki \\ 
\vspace{3.mm}
Institute for Theoretical Physics, Kanazawa University, 
Kanazawa 920-1192, Japan
}

\begin{document}

\begin{abstract}
We study the perfect monopole action as an infrared effective 
theory of SU(2) QCD. It is transformed exactly into a lattice
string model. Since the monopole interactions are weak in the 
infrared SU(2) QCD, the string interactions become 
strong. The strong coupling expansion of string model shows 
the quantum fluctuation is small. The classical string tension 
is estimated analytically, and we see it is very close to the 
quantum one in the SU(2) QCD. We also discuss how to calculate
the glueball mass in our model.
\end{abstract}

\maketitle

\input epsf

{\bf \hspace{-0.35cm}1.$\ $INTRODUCTION}

The infrared effective theory of QCD is important for 
the analytical understanding of  hadron physics. Abelian monopoles 
which appear after
abelian projection of QCD \cite{'thooft} seem to be relevant
dynamical degrees of freedom for infrared region \cite{domi}. 
Shiba and Suzuki \cite{shiba_suzuki} derived the monopole 
action from vacuum configurations obtained in Monte-Carlo simulations
extending the method developed by Swendsen.

We studied the renormalized monopole action ${\cal S}[k]$ performing 
block spin transformations up to $n=8$ numerically, 
and saw that scaling for fixed physical length $b$ looks good\cite{nakam}.
If the action ${\cal S}[k]$ also satisfies the continuum 
rotational invariance, then we can regard ${\cal S}[k]$ as
a good approximation of the renormalized trajectory(RT).
In order to check this, we have to determine the correct form of 
physical operators on the coarse lattice.

\vspace{0.4cm}
{\bf \hspace{-0.35cm}2.$\ $ BLOCKING MONOPOLE CURRENT FROM THE
CONTINUUM AND THE PERFECT OPERATOR}
%

Our strategy is following. (I) Let us start from the following monopole 
action composed of two-point interactions between magnetic monopole 
currents formulated on an infinite lattice with very small lattice 
constant $a$:
\begin{eqnarray}
S[k]=\sum_{s,s',\mu}k_\mu(s)D_0(s-s')k_\mu(s').\label{eqn.S}
\end{eqnarray}
 We have adopted here only parallel interactions, since we can avoid 
perpendicular interactions from short-distance terms using the current 
conservation. 
Moreover, for simplicity, we adopt only the first three Laurent
expansions, i.e., Coulomb, self and nearest-neighbor 
interactions. Explicitly, $D_0(s-s')$ is expressed as 
$\beta\Delta_L^{-1}(s-s')+\alpha\delta_{s,s'}+\gamma\Delta_L(s-s')$.
Here $\Delta_L(s-s')=-\sum_\mu\partial_\mu\partial'_\mu\delta_{s,s'}$ 
and $\partial(\partial')$ is the forward (backward) difference.

The monopole contribution to the potential between
static abelian electric charges is derived from the following
operator\cite{stack,misha1,shiba1}
\begin{eqnarray}
\!\!\!\!\!
W_m({\cal C}) \!\!\!\!\!\! &=& \!\!\!\!\!\! 
 \exp(2\pi i\sum_{s,\mu}N_{\mu}(s,S^J)k_{\mu}(s))
\label{eqn.WC}\\
\!\!\!\!\!
N_{\mu}(s,S_J) \!\!\!\!\!\! &=& \!\!\!\!\!\!
 \sum_{s'}\Delta_L^{-1}(s\! - \! s')\frac{1}{2}
\epsilon_{\mu\alpha\beta\gamma}\partial_{\alpha}
S^J_{\beta\gamma}(s' \!\! + \! \hat{\mu}), \label{eqn.N}
\end{eqnarray}
where $S^J_{\beta\gamma}(s'+\hat{\mu})$ is a plaquette variable satisfying 
$\partial'_{\beta}S^J_{\beta\gamma}(s)=J_{\gamma}(s)$ and the coordinate 
displacement $\hat{\mu}$ is due to the interaction between dual
variables.

(II) Performing a block spin transformation for monopole currents
analytically,
we obtain $\kbra{W_m(C)}$ and the effective action on the coarse 
lattice ($b=n\cdot a$)\cite{fuji99}. 

(III)  Since we see the (numerically obtained) effective monopole 
action for SU(2) QCD in the IR region is well dominated by quadratic 
interactions, we regard the renormalization flow obtained in (II) as 
a projection of RT to the quadratic-interaction plane. We determined 
the couplings in (\ref{eqn.S}) from the monopole action obtained by 
inverse Monte Carlo method.
\begin{figure}
\epsfysize=70mm
 \vspace{-20pt}
 \begin{center}
 \leavevmode
\epsfbox{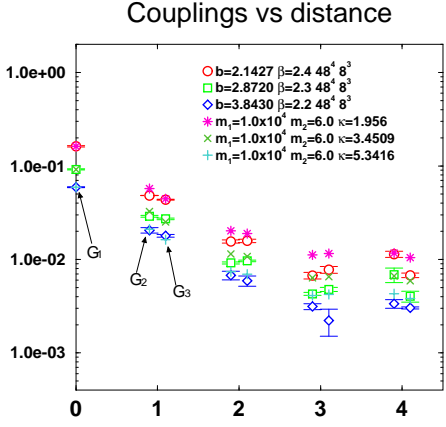}
 \end{center}
 \vspace{-40pt}
{\footnotesize
\caption{
The \(R^2\) dependence of the monopole action
from Swendsen's method and analytical block spin transformation
at $b=2.14, 2.87, 3.84$, where $ \kappa/(m_1^2-m_2^2)=\gamma, 
   m_1^2+m_2^2=\alpha/\gamma, 
   m_1^2m_2^2=\beta/\gamma$.
\label{Mono}
}}
\vspace{-15pt}
\end{figure}
The optimal values $\kappa$, $m_1$ and $m_2$
for $b=$ 2.1, 2.9 and 3.8 (in unit of $\sigma_{phys}^{-1/2}$) 
are illustrated in Figs. 1. 

(IV) The above monopole action can be transformed exactly into 
that of the string model using BKT transformation on the lattice
\cite{misha93,maxim98}. We find finally the correct form of Wilson
loop operator on the coarse lattice as follows:
\begin{eqnarray}
\!\!\!\!\!\!\!\!\!&&
\langle W_m({\cal C}) \rangle \!=\! \nonumber \\
\!\!\!\!\!\!\!\!\!&& 
  \exp
  \Bigg\{\!\!\!
    -\pi^2 \int_{-\infty}^{\infty}\!\!\!\!d^4xd^4y
   N_{\mu}(x)D_0^{-1}(x-y)N_{\mu}(y)
  \Bigg\}\nonumber\\
\!\!\!\!\!\!\!\!\!&&
\times 
  \frac{1}{Z} \sum_{\sigma_{\mu\nu}(s)=-\infty
        \atop{\partial_{[\alpha}\sigma_{\mu\nu]}(s)=0}}^{\infty}
  \!\!\!
  \exp
  \Bigg\{\!
    - {\cal S}[\sigma]
\nonumber \\
\!\!\!\!\!\!\!\!\!&&
    -2\pi^2\sum_{s,s'\atop{\mu,\nu}}\sigma_{\mu\nu}(s)\partial_{\mu}'
    \Delta_L^{-1}(s-s')B_{\nu}(s') 
  \Bigg\},
\label{opwil:5}
\end{eqnarray}
where integer valued two-form $\sigma_{\mu\nu}(s)$ stands for string 
field and ${\cal S}[\sigma]$ is string action\cite{fuji99}. 

\vspace{0.4cm}
{\bf \hspace{-0.35cm}3.$\ $ROTATIONAL INVARIANCE AND THE STRING TENSION}

It turns out that the monopole action on the dual lattice 
is in the weak coupling region for large $b$, namely the infrared region 
of pure $SU(2)$ QCD. Then the string model on the original lattice 
is in the strong coupling region. The strong coupling expansion on 
the lattice can be performed easily and quantum fluctuations terms which 
include more plaquettes become small\cite{kato98}. Thus we consider only
classical part (N-N term) in (\ref{opwil:5}) below.

The plaquette variable $S_{\alpha\beta}$ in Eq.(\ref{eqn.N}) for the 
static potential $V(bI,0,0)$ is expressed by
\begin{eqnarray}
S_{\alpha\beta}(z)
\!&=&\!
  \delta_{\alpha 1}\delta_{\beta 4}\delta(z_{2})\delta(z_{3})
  \theta(z_{1})\theta(Ib\! -\! z_{1}) \nonumber \\
&&\times
  \theta(z_{4})\theta(Tb\! -\! z_{4}).
\label{S1}
\end{eqnarray}

To perform the integration in (4) we go to momentum space.  
Substituting (\ref{S1}) into (4), we get 
\begin{eqnarray}
\!\!\!\!\!\!\!\!&&\!\!\!
{\langle W(Ib,0,0,Tb) \rangle \atop{} }
{\longrightarrow\atop{T\rightarrow\infty\atop{b\rightarrow\infty}}}
\nonumber \\
\!\!\!\!\!\!\!\!&&\!\!\!
\exp\left\{\!\!
   -\pi^2 (TIb^2)\!\!\int\!\!\frac{d^2p}{(2\pi)^2}
   \!\!\left[\frac{1}{\Delta D_0}\right]\!\!(0,p_{2},p_{3},0)    
   \right\}.
\label{eqn.6}
\end{eqnarray}
In (\ref{eqn.6}), since we study large $T$ and large $b$ behaviors, 
we used the following expression:
\begin{eqnarray}
\lim_{T\rightarrow\infty}
\left(\frac{\sin \alpha T}{\alpha}\right)^2
&=& \pi T \delta(\alpha).
\label{pot:1}
\end{eqnarray}

Similarly we can evaluate $\langle W(Ib,Ib,0,Tb) \rangle$.
We choose the variable $S_{\alpha\beta}$ for the static potential
$V(bI,Ib,0)$ as
\begin{eqnarray}
\!\!\!\!\!\!\!\!&&\!\!\! S_{\alpha\beta}(z)
=
  \Bigl(
    \delta_{\alpha 1}\delta_{\beta 4}+\delta_{\alpha 2}\delta_{\beta 4}
  \Bigr)
  \delta(z_{3})\theta(z_{4})\theta(Tb-z_{4})
\nonumber \\
\!\!\!\!\!\!\!\!&&
\times
  \theta(z_{1})\theta(Ib-z_{1})
  \theta(z_{2})\theta(Ib-z_{2})
  \delta(z_{1}-z_{2}).
\end{eqnarray}

$\!\!\!$The static potentials $V(Ib,0,0)$ and $V(Ib,Ib,0)$\\ 
can be written as
\begin{eqnarray}
V(Ib,0,0) \!\!\!\!&=&\!\!\!\!
  \pi^2 (Ib) \int\!\!\frac{d^2p}{(2\pi)^2}\!
  \left[
    \frac{1}{\Delta D_0}
  \right](0,p_2,p_3,0),
\nonumber \\
\!\!\!\!&=&\!\!\!\! \frac{\pi\kappa Ib}{2} \ln\frac{m_1}{m_2} \label{pot1}\\
V(Ib,Ib,0) \!\!\!\!&=&\!\!\!\! 
\frac{\sqrt{2}\pi\kappa Ib}{2} \ln\frac{m_1}{m_2}. 
\end{eqnarray}
The potential takes only the linear form and {\bf the continuum
rotational invariance is recovered completely even for the 
nearest $I=1$ sites}. 
The string tension is evaluated as 
$\sigma_{cl}=\frac{\pi\kappa}{2} \ln\frac{m_1}{m_2}$. 
This is consistent with the analytical results\cite{suzu89}.
The two constants $m_1$ and $m_2$ can be regarded as the coherence and
the penetration lengths.

The ratio $\sqrt{\sigma_{cl}/\sigma_{phys}}$ from the optimal values 
$\kappa$, $m_1$ and $m_2$ become 1.64, 1.56 and 1.45 for $b=$ 2.1, 2.9
and 3.8, respectively.
The scaling for physical length $b$ seems to be good, although $\sigma_{cl}$
deviate a little from physical string tension $\sigma_{phys}$.

We can also evaluate the string tension from the large 
flat Wilson loop not via $D_0$ by following expression:
\begin{eqnarray}
\!\!\!\!\!\!&&\sigma = \int_{-\pi}^{\pi}\frac{d^2p}{(2\pi)^2}
  \Delta_{\rm L}^{-2}(k_1,k_2,0,0) 
\nonumber \\
\!\!\!\!\!\!&&\times
  \left[
   \sin^2\frac{k_2}{2}D^{-1}(k_1,k_2,0,0;\hat{1}) 
  + (1 \leftrightarrow 2)
  \right].
\end{eqnarray}
We can show this is equivalent to the above formula (\ref{pot1}). 
In this case $\sqrt{\sigma_{cl}/\sigma_{phys}}$ become 
1.73, 1.59 and 1.39 for $b=$ 2.1, 2.9
and 3.8, respectively.
From these results, we see the discrepancy does not
come from the systematic error of fitting $D_0$.
It may come from the fact that
we use the definition of the {\bf monopole a la DeGrand} 
not but the {\bf real monopole} in numerical study
\footnote{The real monopole is a monopole whose magnetic charge 
run from $-\infty$ to $\infty$, while the charge of the
monopole a la DeGrand run from $-(3n^2-1)$ to $(3n^2-1)$ by
definition.\cite{D_T,ivanenko}}.

\vspace{0.4cm}
{\bf \hspace{-0.35cm}4.$\ $ON THE GLUEBALL MASS}

The glueball mass spectrum can be obtained by computing the
correlation functions of gauge invarient local operators or 
Wilson loops, and extract the particle poles. For examples,
one can consider two point function of operator 
${\cal O}= Tr(F^2)$. For large $|x-y|$ it can be expanded as
\begin{eqnarray}
 <{\cal O}(x){\cal O}(y)> \simeq \sum c_i \exp(-M_i|x-y|),
\end{eqnarray}
where $M_i$ are called the glueball mass.

In our model, we take the operator ${\cal O}$ as
\begin{eqnarray}
 {\cal O}= \frac{1}{a^4}(1-W_m({\cal C}))
\end{eqnarray}
on the $a$-lattice. We can show this operator coinside with 
abelian counter part of ${\cal O}= Tr(F^2)$ when lattice constant
$a$ goes to zero. The evaluation of its mass is now in progress.

\vspace{0.4cm}
{\bf \hspace{-0.35cm}5.$\ $CONCLUSIONS}

We found the {\bf quantum perfect lattice action} for monopoles and
strings of hadron which {\bf describe low-energy SU(2) QCD} by using
analytical blockspin transformation and numerical simulation.
{\bf The strong coupling expansion on the lattice} for the hadronic
string model works good in the infrared region of SU(2) QCD
and yields the string tension. 
The results are {\bf almost consistent with the recent lattice results}.
The discrepancy may come from the definition of the 
{\bf monopole a la DeGrand}.


\end{document}